\documentclass[a4paper,
               biblatex,     
               ]{jacow}
\usepackage{pdfpages,multirow,ragged2e} %
\makeatletter%
	\ifboolexpr{bool{xetex}}
	 {\renewcommand{\Gin@extensions}{.pdf,%
	                    .png,.jpg,.bmp,.pict,.tif,.psd,.mac,.sga,.tga,.gif,%
	                    .eps,.ps,%
	                    }}{}
\makeatother

\ifboolexpr{bool{xetex} or bool{luatex}} 
 {}                                      
 {\usepackage[utf8]{inputenc}}           

\usepackage[USenglish]{babel}

\ifboolexpr{bool{jacowbiblatex}}%
 {%
  \addbibresource{MOPS75.bib}
 }{}
\begin{document}

\title{Towards latent space evolution of spatiotemporal dynamics of six-dimensional phase space of charged particle beams \thanks{Work supported by the LANL LDRD Program Directed Research (DR) project 20220074DR}}

\author{M. Rautela \thanks{mrautela@lanl.gov}\textsuperscript{1}, A. Williams\textsuperscript{1,2}, A. Scheinker\textsuperscript{1} \\ 
        \textsuperscript{1}Los Alamos National Laboratory, NM, US \\
        \textsuperscript{2}University of California, San Diago, CA, US}
	
\maketitle

\begin{abstract}
Addressing the charged particle beam diagnostics in accelerators poses a formidable challenge, demanding high-fidelity simulations in limited computational time. Machine learning (ML) based surrogate models have emerged as a promising tool for non-invasive charged particle beam diagnostics. Trained ML models can make predictions much faster than computationally expensive physics simulations. In this work, we have proposed a temporally structured variational autoencoder model to autoregressively forecast the spatiotemporal dynamics of the 15 unique 2D projections of 6D phase space of charged particle beam as it travels through the LANSCE linear accelerator. In the model, VAE embeds the phase space projections into a lower dimensional latent space. A long-short-term memory network then learns the temporal correlations in the latent space. The trained network can evolve the phase space projections across further modules provided the first few modules as inputs. The model predicts all the projections across different modules with low mean squared error and high structural similarity index.
\end{abstract}

\section{INTRODUCTION}
With the advancement in parallel processing, machine learning (ML) and deep learning (DL) have shown promising capabilities in solving problems in physics. Most of the problems in physics are governed by spatiotemporal dynamics, where complex spatial behavior evolves with time \cite{rautela2023bayesian}. In a particle accelerator, the dynamics of charged particles evolve temporally in a six-dimensional phase space made up of three position and three momentum components for each particle i.e., $(x, y, z, p_x, p_y, p_z)$ with $z$ typically chosen as the direction along the accelerator axis \cite{scheinker2020adaptive_JAP}.

The majority of the ML research is inclined towards learning spatial or temporal dynamics with limited emphasis on spatiotemporal dynamics. Some of the DL techniques for solving spatiotemporal dynamical problems are three-dimensional convolutional neural networks (3DCNN) \cite{wandel2021teaching}, convolutional long short-term memory (ConvLSTM) \cite{shi2015convolutional}, Deep Convolutional Generative Adversarial Networks (DCGAN) \cite{cheng2020data}, Graph Neural Networks (GNNs) \cite{kipf2016semi}. 

Recently, latent evolution models have gained traction for solving spatiotemporal dynamics problems. In these computationally efficient models, a dimensionality reduction framework learns spatial correlations by mapping higher dimensional images to lower dimensional latent space. For example, in Ref.~\cite{scheinker2021adaptive_JOI}, an adaptive virtual 6D phase space diagnostic was developed for particle accelerator beams, where an autoencoder maps high dimensional images representing the states of complex time-varying particle accelerator beams to a low-dimensional latent space from which it then generates all 15 unique 2D projections of the beam's 6D phase space. Adaptive feedback is used within the low-dimensional latent representation to track an unknown time-varying beam's properties with time as the accelerator parameters change. A robustness study of this adaptive latent space tuning method has demonstrated an ability to extrapolate beyond the span of the training data with a more physically consistent generated 6D phase space \cite{scheinker2023adaptive,scheinker2021adaptive_SciRep}. A combination of autoencoders for learning spatial dynamics and LSTMs for temporal dynamics has also been employed for addressing fluid flow problems \cite{wiewel2019latent, nakamura2021convolutional, maulik2021reduced, vlachas2022multiscale}.

In this paper, we introduce a two-step deep learning modeling framework, wherein we study the full spatiotemporal dynamical nature of the evolution of a charged particle beam through various sections of a linear accelerator. In this model, a conditional variational autoencoder (CVAE) is used to learn a low-dimensional latent space distribution of the high-dimensional phase space of charged particles. An LSTM-based recurrent neural network is employed to learn the temporal dynamics within the latent space. It gives the model two promising abilities, allowing both the generation of realistic projections across different modules and the forecasting of phase space in further modules \cite{rautela2024conditional}.

\section{METHODS}
\subsection{Multi-Particle Tracking Simulations}
The behavior of the charged particles in accelerators is governed by numerous accelerator parameters like radio frequency cavity field, and magnetic field strengths. During accelerator operation, these parameters are manually adjusted to minimize beam loss. However, this manual adjustment process is time-consuming and often results in suboptimal performance. Moreover, the time-varying nature of these parameters introduces further uncertainties. Non-destructive beam measurements are scarce in most accelerators, primarily due to resolution limits with short-duration beam pulses and short run times, posing challenges to meaningful data collection. To achieve optimal functionality, understanding beam dynamics is crucial. 

Various simulation tools have been devised for this purpose \cite{tenenbaum2005lucretia,young2003particle,pang2014gpu}. High-Performance Simulator (HPSim) is an advanced multiple-particle beam dynamics simulator taking into account the effects of external accelerating and focusing forces as well as space charge forces \cite{pang2014gpu}. In this work, we generate synthetic data by randomly sampling RF set points (amplitude and phase) of the first four modules from a uniform distribution keeping other beam and accelerator parameters fixed. This investigation centers on the LANSCE linear accelerator at Los Alamos National Laboratory \cite{wangler2008rf}. More details about the optimization and tuning challenges of the LANSCE accelerator are given in Ref.~\cite{scheinker2021extremum_LANSCE}.

\begin{figure}
    \centering
    \includegraphics[width=0.5\textwidth]{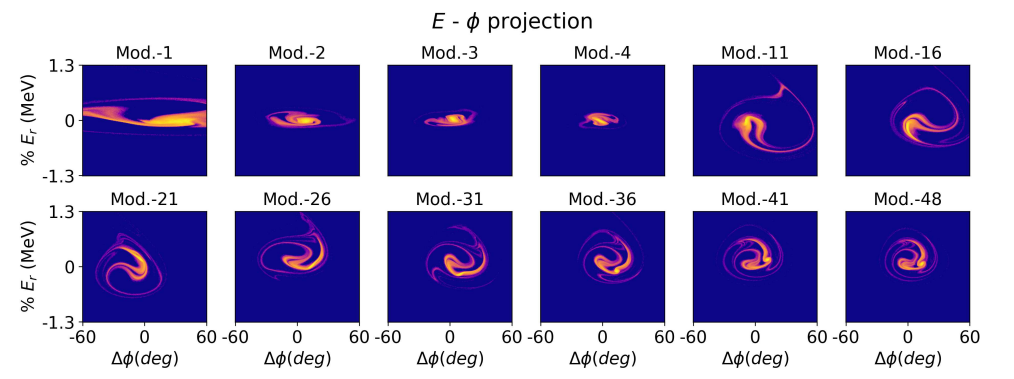}
    \caption{$E-\phi$ projection at various accelerating modules. The images are unnormalized and shown on a logarithmic scale.}
    \label{fig:dataset}
\end{figure}

From the HPSim output, we generate 2D histograms which are the 15 unique projections of the beam's 6D $(x,y,z,p_x,p_y,p_z)$ phase space at each of the 48 modules. For our application $(z,p_z)$ is converted to $(\phi,E)$ where $\phi$ is the phase of a particle in a bunch relative to the design phase. In Fig.~\ref{fig:dataset}, $E,\phi$ projection is plotted at various accelerating modules. The plots are shown on a logarithmic scale for better visualization.

\subsection{Latent Space Evolution via CVAE-LSTM}
Autoencoders (AE) are able to learn low-dimensional latent representations of complex data and then generate new high-dimensional data from the latent embedding \cite{rautela2022delamination}. Variational autoencoders (VAE) are AEs that map to a probabilistic latent space. Due to this, VAEs enable the generation of new realistic samples by traversing the latent space \cite{rautela2022towards,rautela2023deep}. In this work, we propose a CVAE-LSTM-based latent evolution model that utilizes a conditional VAE to transform 6D phase space into a lower dimensional latent space. An LSTM-based recurrent neural network learns the temporal dynamics within the latent space. The uniqueness of the model lies in its ability to independently learn spatial and temporal dynamics through a two-step process. 

\begin{figure}[h!]
    \centering
    \includegraphics[width=0.5\textwidth]{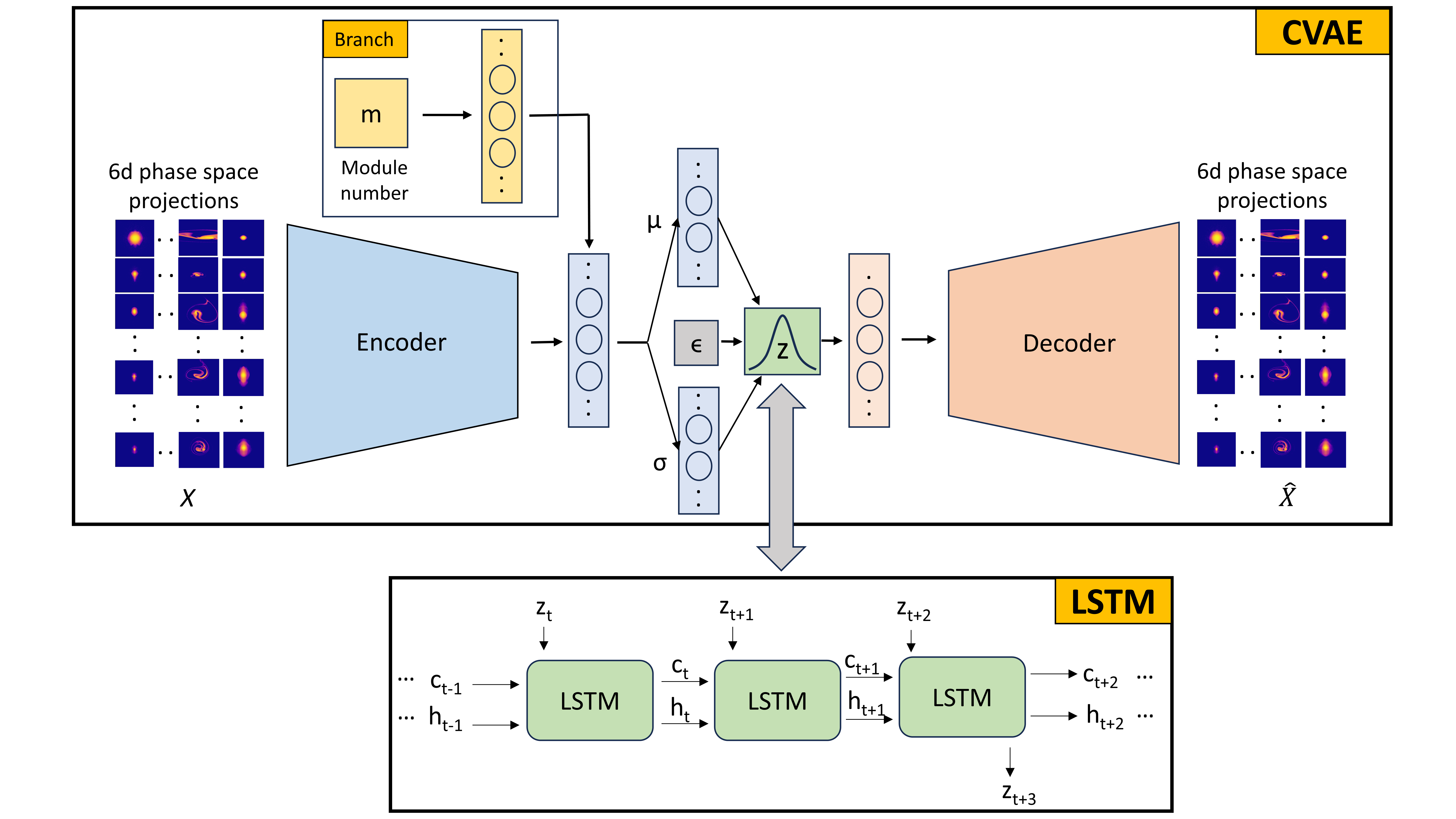}
    \caption{CVAE-LSTM as a latent evolution model. CVAE maps phase space projections to latent space and a LSTM learns to forecast future downstream states autoregressively given previous upstream states.}
    \label{fig:CLARM}
\end{figure}

The architecture of the CVAE-LSTM is represented in Fig.~\ref{fig:CLARM}. The initial 15 unique phase space projections are depicted as 15 channels, each represented by a 256 $\times$ 256-pixel image, resulting in a $\sim10^6$ dimensional input that is encoded into an 8-dimensional latent space. The projections along with the module number (1-48) are input to the CVAE. The encoder performs convolution operations, and the extracted features are concatenated with the module number and then input to the CVAE's latent space. The learned latent space is then processed by an LSTM network, which is designed to forecast the next latent space point (corresponding to projections in the next module) based on previous points (corresponding to projections in the previous modules). The continuous latent space of the CVAE allows for conditional sampling, followed by a decoder which generates realistic projections across different modules of the accelerator.

\section{RESULTS}
To generate data, the RF field set points (amplitude and phase) of the DTL sections (first 4 modules) were randomly sampled (1400 simulations) while keeping the other 88 RF parameters fixed. HPSim would then simulate the dynamics of a beam through the entire accelerator, from which the 15 unique phase space projections (each with a 256 $\times$ 256 image) at each of the 48 RF modules were generated. 1400 simulation data sets were used for training and 100 for testing. A single input to the VAE is a set of 15 256 $\times$ 256-pixel images. The conditional input $c$ to the encoder is the module number, a scalar between 1-48, normalized to the range $[0,1]$.  

\subsection{Latent Space Visualization}
\begin{figure}[h!]
    \centering
    \includegraphics[width=0.5\textwidth]{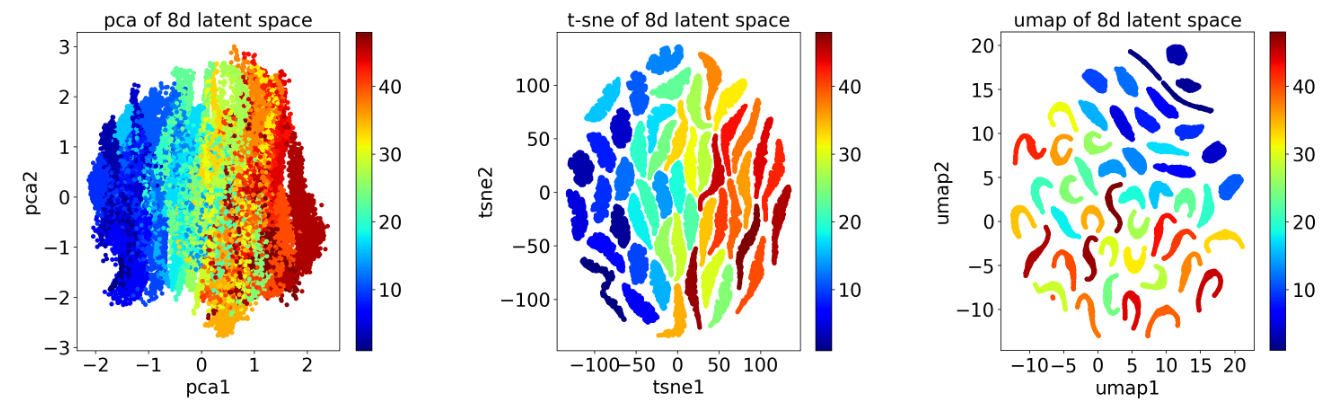}
    \caption{2D PCA, t-SNE and UMAP of 8D latent space. The color maps corresponding to different module number.}
    \vskip-5mm
    \label{fig:latentvisualization}
\end{figure}

Visualization of the latent space is important for the interpretability of the network. However, the visualization is restricted by the high dimensionality (8D) of the latent space. We have transformed the 8D latent space is transformed into various different 2D spaces, as shown in Fig.~\ref{fig:latentvisualization}. The first method is a linear dimensionality reduction technique called principal component analysis (PCA). The other two methods are manifold learning techniques called t-distributed Stochastic Neighbor Embedding (t-SNE) \cite{van2008visualizing} and Uniform Manifold Approximation and Projection (UMAP) \cite{mcinnes2018umap}. Both of them are non-linear dimensionality reduction approaches, contrary to PCA. While t-SNE is a more popular method for various problems, UMAP performs better in preserving both local and global structures, computational efficiency, and parameter robustness \cite{mcinnes2018umap}.

\subsection{Forecasting Ability}
\begin{figure}[h!]
    \centering
    \begin{minipage}[b]{0.49\linewidth}
        \centering
        \includegraphics[width=1.0\textwidth]{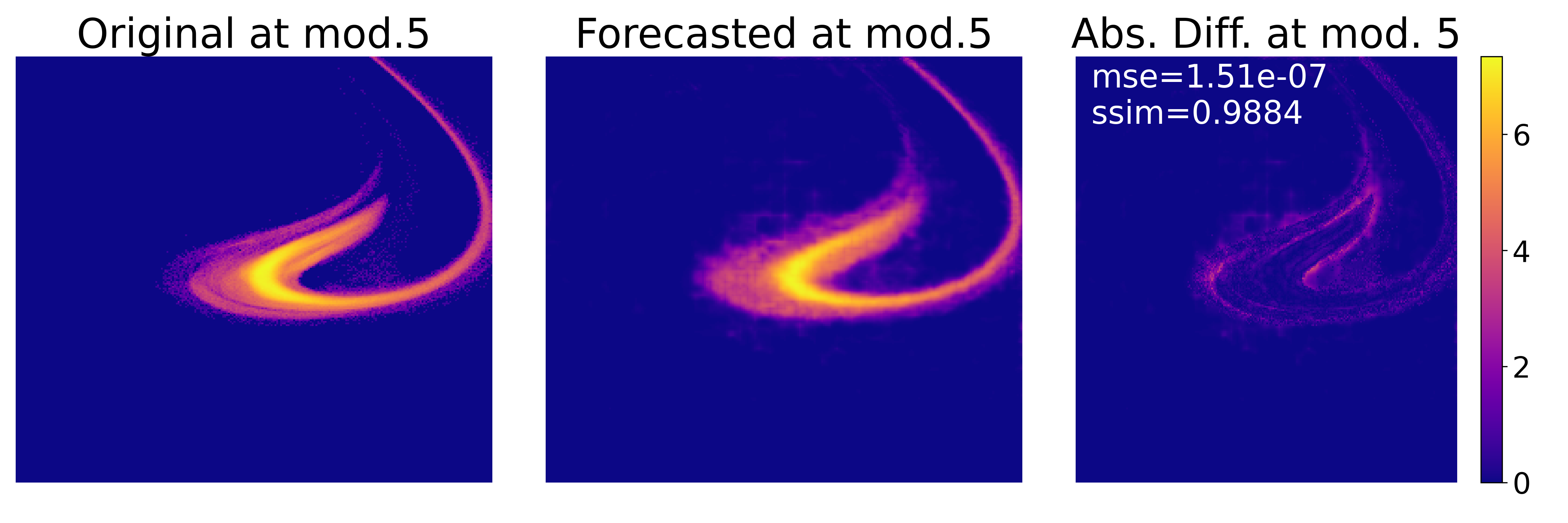}
    \end{minipage}
    \begin{minipage}[b]{0.49\linewidth}
        \centering
        \includegraphics[width=1.0\textwidth]{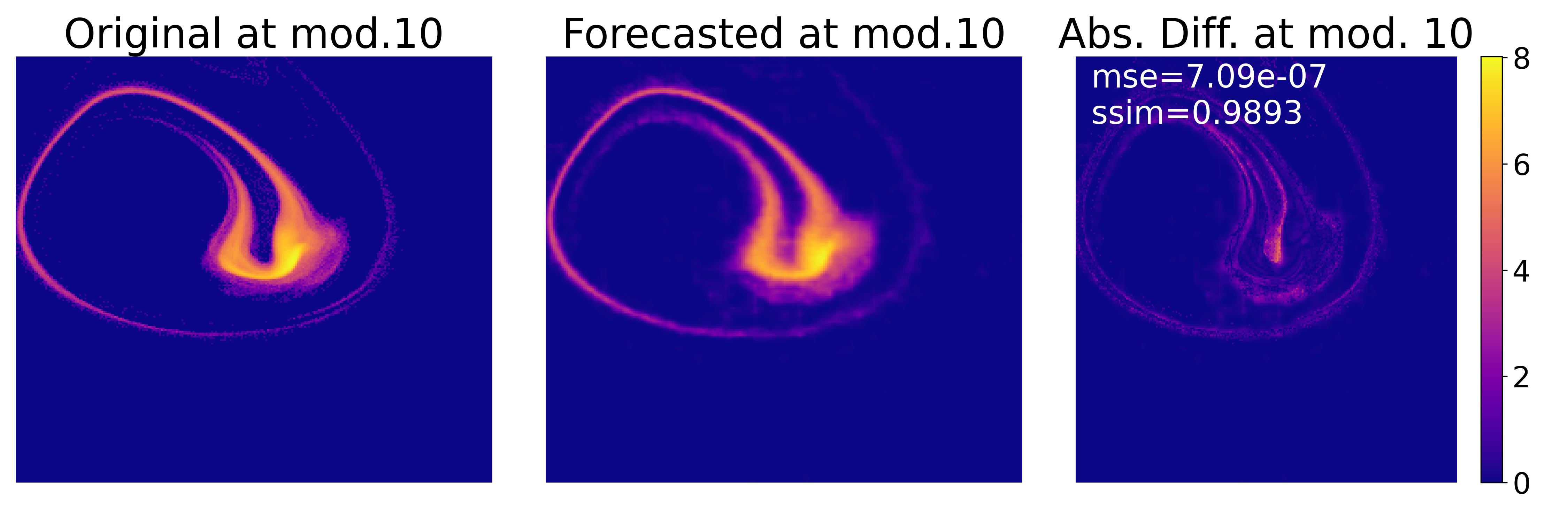}
    \end{minipage}
    \begin{minipage}[b]{0.49\linewidth}
        \centering
        \includegraphics[width=1.0\textwidth]{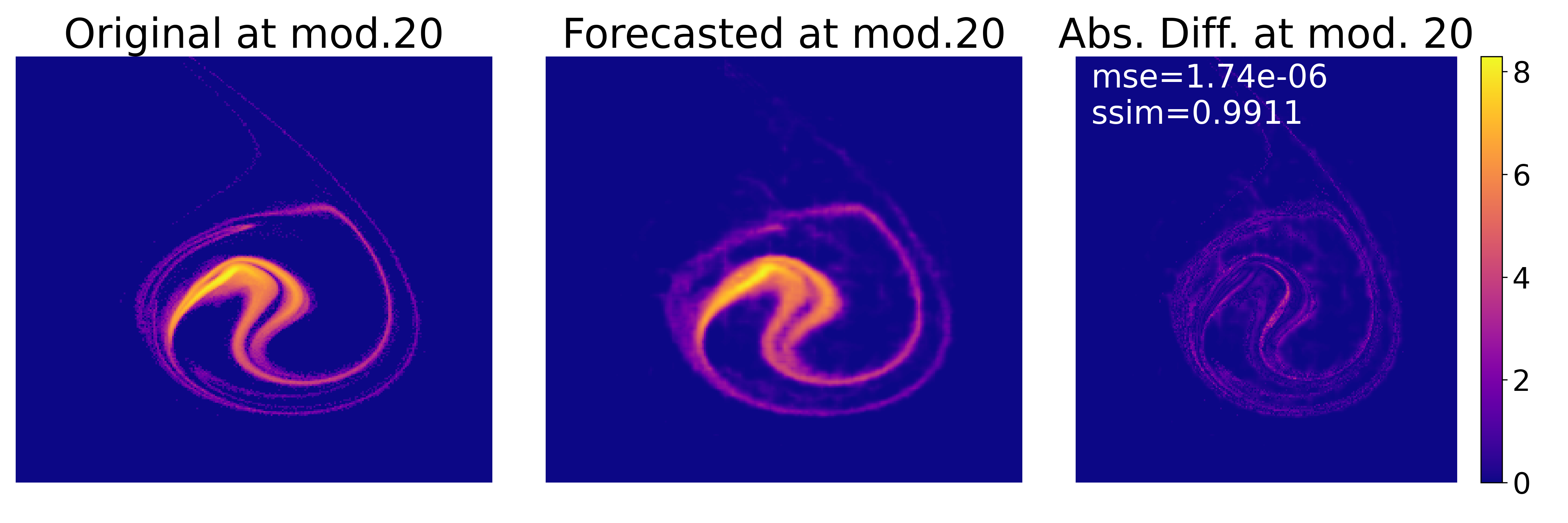}
    \end{minipage}
    \begin{minipage}[b]{0.49\linewidth}
        \centering
            \includegraphics[width=1.0\textwidth]{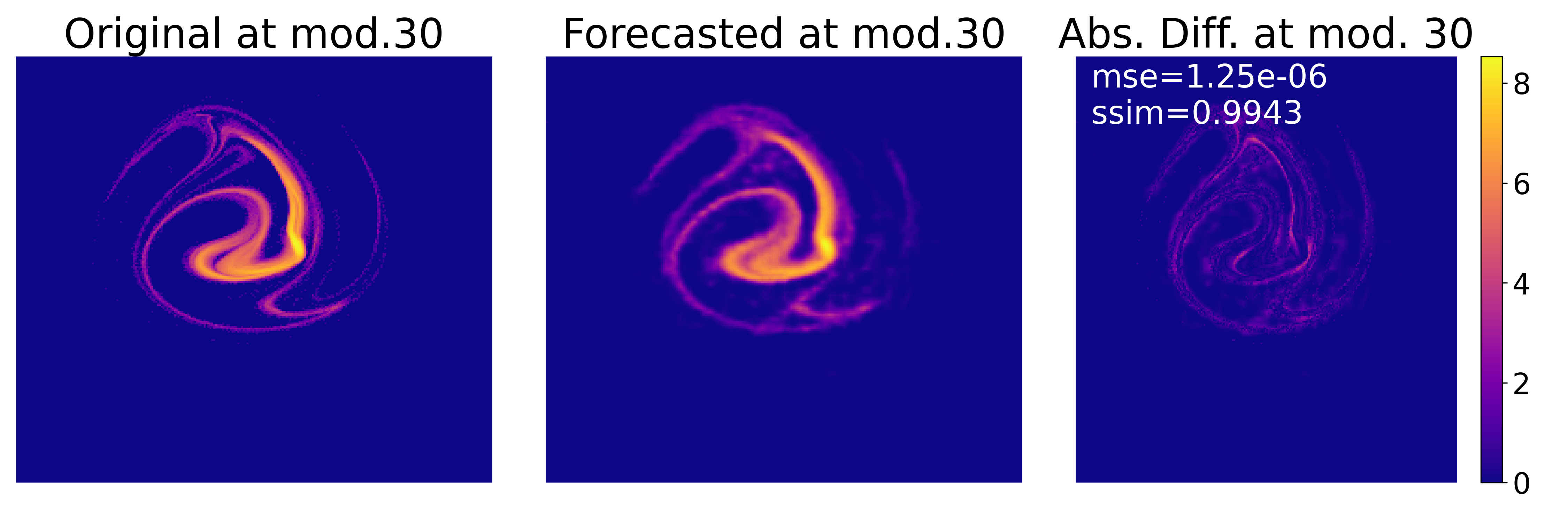}
    \end{minipage}
    \begin{minipage}[b]{0.49\linewidth}
        \centering
        \includegraphics[width=1.0\textwidth]{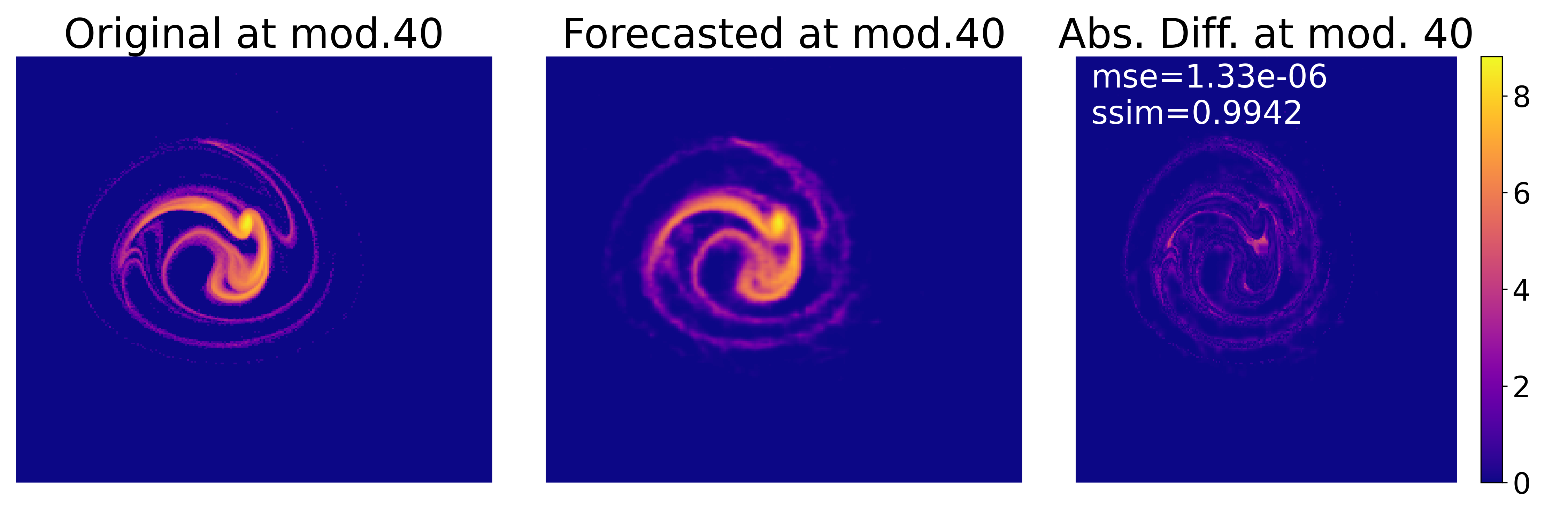}
    \end{minipage}
    \begin{minipage}[b]{0.49\linewidth}
        \centering
        \includegraphics[width=1.0\textwidth]{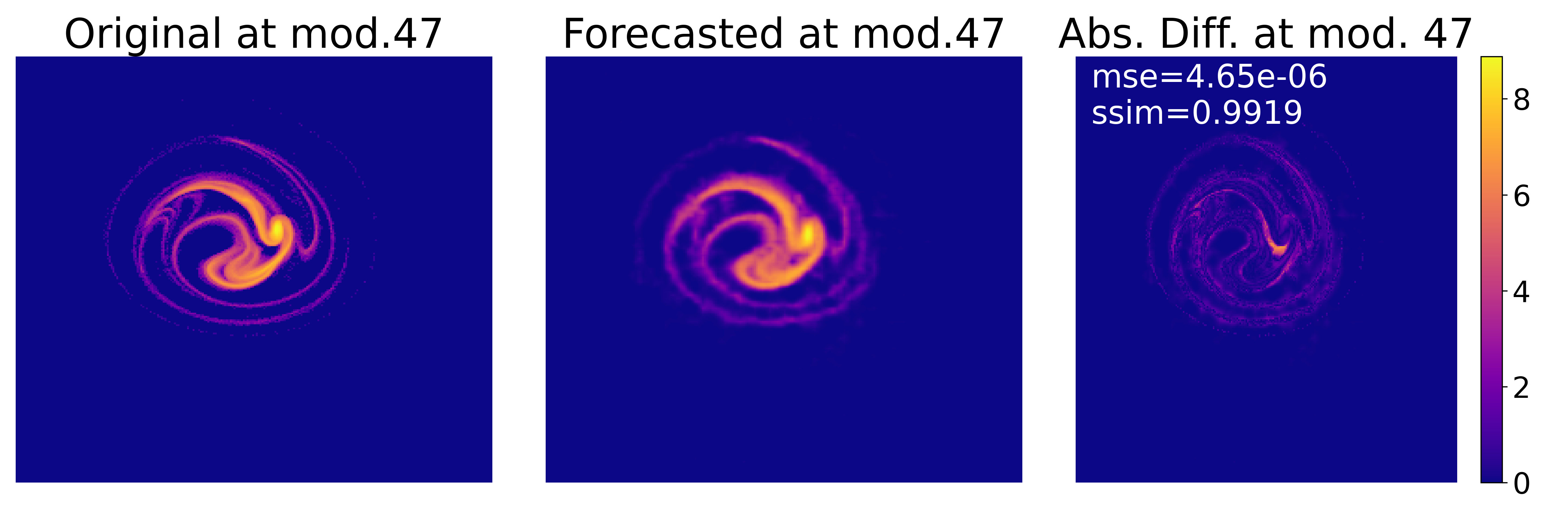}
    \end{minipage}
    \caption{Forecasting results: Forecasted projections (shown $E-\phi$ only) across different modules given first four projections as inputs. The original projection is presented against the forecasted along with the absolute difference between both.}
    \label{fig:forecastingability}
\end{figure}

The test set is projected onto the latent space and LSTM is used to forecast downstream latent points given upstream latent points. The forecasted latent points are passed through the decoder of the CVAE to reconstruct the phase space projections. In Fig.~\ref{fig:forecastingability}, we showcase the $E-\phi$ projection using the initial four-phase space projections from the test set as input. The figure depicts the original projection, the predicted projection, and their absolute difference at various modules. The corresponding MSE and SSIM for the projection reveal high similarity, with MSE of the order of $10^{-7}$ and SSIM exceeding 0.99. We have noted that the discrepancies in forecasted projections grow for later modules, particularly MSE. This growth in MSE is expected as the LSTM's inputs are the true values of $M_1-M_4$, based on which it predicts an estimate $\hat{M_5}$ of $M_5$ and then uses its own prediction to generate $\hat{M_6}$ and so on in an iterative manner in which the errors introduced by the CVAE and LSTM are propagated and accumulated leading to a continuous increase in error.

The forecasting of all the projections in all the modules takes less than one second whereas HPSim takes around 10~minutes with similar computing infrastructure, resulting in a speed up by a factor of $\sim 600$. The exceptional computational speed of the method makes it extremely well-suited for various real-time accelerator applications. The method can be used as a virtual diagnostic in which CVAE-LSTM predicts a detailed evolution of the beam's phase space through the entire LANSCE accelerator based on the current RF module settings and using only 4 initial steps from the much slower HPSim physics-based model as its initial points. In general, the application of such an approach to any large accelerator will provide a substantial benefit for simulating beam dynamics and for accelerator optimization.

Uncertainty analysis is a byproduct of probabilistic models (like VAE) and it plays an important role in understanding uncertainties associated with the accelerator operation. In our proposed methods, just by sampling the latent space for the first few modules, the LSTM and decoder can be used to generate phase space projections in all the modules. A detailed investigation of the uncertainty analysis aspect is a part of future research work.

\section{CONCLUSION}
A novel latent evolution model i.e., CVAE-LSTM is proposed for learning spatiotemporal dynamics. The application of the model is shown for forecasting complex dynamics of charged particle beams through linear accelerators without any supervision from RF field set points. The forecasting results are promising when tested with different evaluation metrics. We have performed a visualization of the latent space PCA, t-SNE, and UMAP. The proposed methodology brings computational speed, robustness, and enhanced interpretability for solving spatiotemporal dynamics problems. This general method provides a computational speed-up of approximately 600x for complex beam dynamics and is applicable to a wide range of accelerator tuning, optimization, and virtual diagnostics applications. 

\section{ACKNOWLEDGEMENTS}
This work was supported by the Los Alamos National Laboratory LDRD Program Directed Research (DR) project 20220074DR.

\ifboolexpr{bool{jacowbiblatex}}%
	{\printbibliography}%
	{%
	
	
}

\end{document}